\iftrue
\documentclass[aps,prl,twocolumn,
			   groupedaddress,superscriptaddress,
			   amsfonts,amssymb,amsmath,
			   citeautoscript,
			   a4paper]{revtex4-1}
\else
\documentclass[aps,pra,preprint,
			   groupedaddress,superscriptaddress,notitlepage,
			   amsfonts,amssymb,amsmath,
			   citeautoscript,
			   a4paper]{revtex4-1}
\fi

\usepackage[pdftex]{hyperref}
\hypersetup{colorlinks,
			linkcolor={blue!75!black!80!yellow},
			citecolor={blue!75!black!80!yellow},
			urlcolor={blue!75!black!80!yellow},
			pdfstartview=FitH}
\usepackage{graphicx}
\usepackage{mathrsfs}
\usepackage{amsthm}
\usepackage{physics}
\usepackage{xspace}
\usepackage{braket}
\usepackage{xr}
\usepackage{gensymb} 
\usepackage{xcolor,soul}
\usepackage{stmaryrd} 
\usepackage[UKenglish]{babel}
\usepackage{placeins} 
\usepackage{siunitx}
\DeclareSIUnit[number-unit-product=]\percent{\char`\%} 
\usepackage{makecell}
\usepackage{diagbox}
\usepackage{float}
\usepackage{amsmath} 
\usepackage{empheq} 

\usepackage{txfonts}  
\usepackage{txfontsb} 

\newcommand{\iu}{\mathrm{i}}
\newcommand{\e}{\mathrm{e}}

\newcommand{\appropto}{\mathrel{\vcenter{
  \offinterlineskip\halign{\hfil$##$\cr
    \propto\cr\noalign{\kern2pt}\sim\cr\noalign{\kern-2pt}}}}}

\newcommand{\ie}{i.e.\@\xspace}  

\newcommand{\eg}{e.g.\@\xspace}

\makeatletter
\newcommand*{\addFileDependency}[1]{
  \typeout{(#1)}
  \@addtofilelist{#1}
  \IfFileExists{#1}{}{\typeout{No file #1.}}
}
\makeatother

\usepackage{scalerel}

\usepackage{textcomp} 
\usepackage{xifthen}
\usepackage{etoolbox}
\newboolean{togglecomments}
\newboolean{togglechanges} 

\setboolean{togglecomments}{false}
\setboolean{togglechanges}{true}

\newcommand{\comment}[2]{%
    \ifbool{togglecomments}%
    {\textcolor{blue!70!black}{\small\textsf{%
    \textsuperscript{\textsc{\textsf{\MakeLowercase{#1}}}}%
    [#2]}}} 
    {}}     
\newcommand{\swap}[2]{\ifbool{togglechanges}
    {#2}  
    {\textcolor{red!70!black}{[#1]}\textrightarrow{}\textcolor{green!50!black}{[#2]}}}
\newcommand{\remove}[1]{\ifbool{togglechanges}
    {}    
    {\textcolor{red!70!black}{#1}}}
\newcommand{\inset}[1]{\ifbool{togglechanges}
    {#1}  
    {\textcolor{green!50!black}{#1}}}
\newcommand{\optional}[1]{\ifbool{togglechanges}
    {#1}  
    {\textcolor{yellow!50!orange!80!gray}{#1}}}
\newcommand{\citeremind}[1]{%
    [\textcolor{blue!75!black!80!yellow}{
        $\blacksquare$%
           \ifthenelse{\isempty{#1}}
               {}
               {\textsuperscript{\textsf{#1}}}%
        }]\xspace}
\newcommand{\todo}[1]{
    \textcolor{orange!80!yellow!95!black}{\textbf{[}%
        \ifthenelse{\isempty{#1}}%
        {\text{$\blacksquare$}}%
        {{\small\textsf{#1}}}%
        \textbf{]}}}
        
\begin{document}

\preprint{APS/123-QED}

\title{Topological quantum electrodynamics in synthetic non-Abelian gauge fields}

\author{Qinan Huang}%
\author{Bengy~T.~T. Wong}%
\author{Zehai Pang}%
\author{Xudong Zhang}%
\author{Zeling Chen}%
\author{Yi Yang}
\email{yiyg@hku.hk}
\affiliation{%
 Department of Physics and HK Institute of Quantum Science and Technology, \\
The University of Hong Kong, Pokfulam, Hong Kong, China
}%

\date{\today}

\begin{abstract}

Quantum electrodynamics (QED), a cornerstone framework that describes light-matter interactions rooted in Abelian symmetries, renders the harnessing of synthetic non-Abelian gauge fields as a fundamental yet uncharted frontier.
Here, we develop a general theory of light-matter interaction of quantum emitters embedded in non-Abelian photonic lattices. 
Based on analytical solutions to the non-Abelian Landau dressed states beyond the continuum limit, we reveal chiral photon emission and vortices with emergent nonreciprocity enabled by selective coupling between emitters and spin-momentum-locked bands.
When coexisting with Abelian and non-Abelian magnetic fields, emitters hybridize with Landau dressed orbits to form spin-polarized, squeezed Landau polaritons that carry quantized angular momenta, with Rabi frequencies tunable via Landau levels and pseudospin interactions. 
Multi-emitter dynamics further exhibit collective phenomena governed by real-space staggered phases induced by nonsymmorphic crystalline symmetry. 
These results bridge non-Abelian physics with quantum optics, and establish non-Abelian gauge fields as a versatile tool for synthesizing topological quantum optical states, angular momentum transfer, and controlling photon-mediated correlations in QED systems, relevant for applications in quantum simulations and chiral quantum optical networks.

\end{abstract}

\maketitle

Synthetic gauge fields provide a powerful platform for exploring and manipulating topological phenomena in atomic, molecular, and optical systems, with non-Abelian gauge fields offering particularly rich opportunities due to their non-commutative nature. These non-Abelian fields can be encoded in photonic modes~\cite{yang2024non,hafezi2011robust,terccas2014non,whittaker2021optical,yang2019synthesis,chen2019non,cheng2023artificial,pang2024synthetic,pang2024topological,cheng2025non} and atomic states~\cite{dalibard2011colloquium,goldman2014light,aidelsburger2018artificial}, allow the manipulation of non-Abelian geometric phases~\cite{wilczek1984appearance,sugawa2021wilson} and synthetic spin-orbit coupling (SOC)~\cite{galitski2013spin,zhai2015degenerate}, and enable a variety of effects like Zitterbewegung ~\cite{merkl2008atomic,vaishnav2008observing,zawadzki2011zitterbewegung,sedov2018zitterbewegung, hasan2022wave, polimeno2021experimental, lovett2023observation,wong2025synthetic}, non-Abelian monopoles~\cite{sugawa2018second}, and non-Abelian Aharonov–Bohm interference~\cite{horvathy1986non,chen2019non,yang2019synthesis,liang2024chiral}. 
In the many-body regime, they are predicted to be useful in creating fractional quantum Hall states and non-Abelian anyonic quasiparticles~\cite{burrello2010non,burrello2011ultracold,grass2013fractional,palmer2011fractional}, while also facilitating the quantum simulations of high-energy phenomena and lattice gauge theories~\cite{zohar2015quantum,halimeh2025cold,tagliacozzo2013simulation,mezzacapo2015non}. 
These prospects indicate the potential of non-Abelian gauge fields for advancing light-matter interaction studies. However, their integration into quantum electrodynamic frameworks---particularly in topological contexts---remains an open question.

This gap now becomes more pertinent through the rapid progress in engineering photon-emitter interfaces within tailored photonic environments ~\cite{lodahl2015interfacing,chang2018colloquium, sheremet2023waveguide, gonzalez2024light}, which has unlocked unprecedented control over light-matter interactions~\cite{lodahl2017chiral, xia2014reversible, picardi2018janus, scheucher2016quantum, sollner2015deterministic, pucher2022atomic, sayrin2015nanophotonic,reitz2022cooperative}. 
A pivotal development in this field is the recognition of the critical role played by the lattice geometry and topological properties of photonic reservoirs in shaping light-matter coupling~\cite{gonzalez2017markovian,gonzalez2018exotic, garcia2020tunable, perczel2020theory, wang2022unconventional,bello2019unconventional,leonforte2021vacancy,sanchez2020chiral,bello2023topological}. For instance, emitters detuned at photonic band edges or into bandgaps exhibit non-Markovian dynamics mediated by conical degeneracies, Van Hove singularities, and bound photon modes~\cite{douglas2015quantum,calajo2016atom,shi2016bound,sanchez2017dynamical}, while multi-emitter systems manifest collective sub- and superradiant states through photon-mediated interference~\cite{gonzalez2017markovian,goban2015superradiance,ramos2016non}.
These advances underscore the capacity of structured photonic systems to host complex gauge field effects, yet most studies have focused on Abelian paradigms.
Under uniform Abelian magnetic fields, edge-localized or edge-coupled emitters interact with chiral edge states, enabling directional decay~\cite{barik2018topological,longhi2019quantum,giorgi2020topological,blanco2018topological,dai2022topologically}, whereas bulk emitters hybridize with photonic Landau orbits to form Landau polaritons—exhibiting Rabi oscillations intertwined with quantized photon dynamics~\cite{de2021light,de2023chiral,li2023light}. However, the exploration of non-Abelian gauge fields in such hybrid emitter-photon systems remains nascent despite their potential to account for the inherently anisotropic and multimode nature of emitter-photon reservoir coupling.

Here, we bridge this gap by exploring emitter-photon interactions on a two-dimensional (2D) photonic lattice immersed in synthetic non-Abelian gauge fields. The synthetic spin-orbit coupling for photons enables the emitters to excite phase vortices amid nontrivial real-space photonic spin textures with emergent non-reciprocity. This idea is further extended to the coexistence of Abelian and non-Abelian fields, where emitters excite squeezed spin-polarized Landau polaritons of various angular momenta whose Rabi frequencies can be tuned by the Landau levels and photonic pseudospins. In the multi-emitter scenario, their dynamics are shown to be strongly shaped by the nonsymmorphic crystalline symmetries of the lattice.

\section{Quantum emitters immersed in non-Abelian gauge fields: Model}

\begin{figure}
\includegraphics[width=\linewidth]{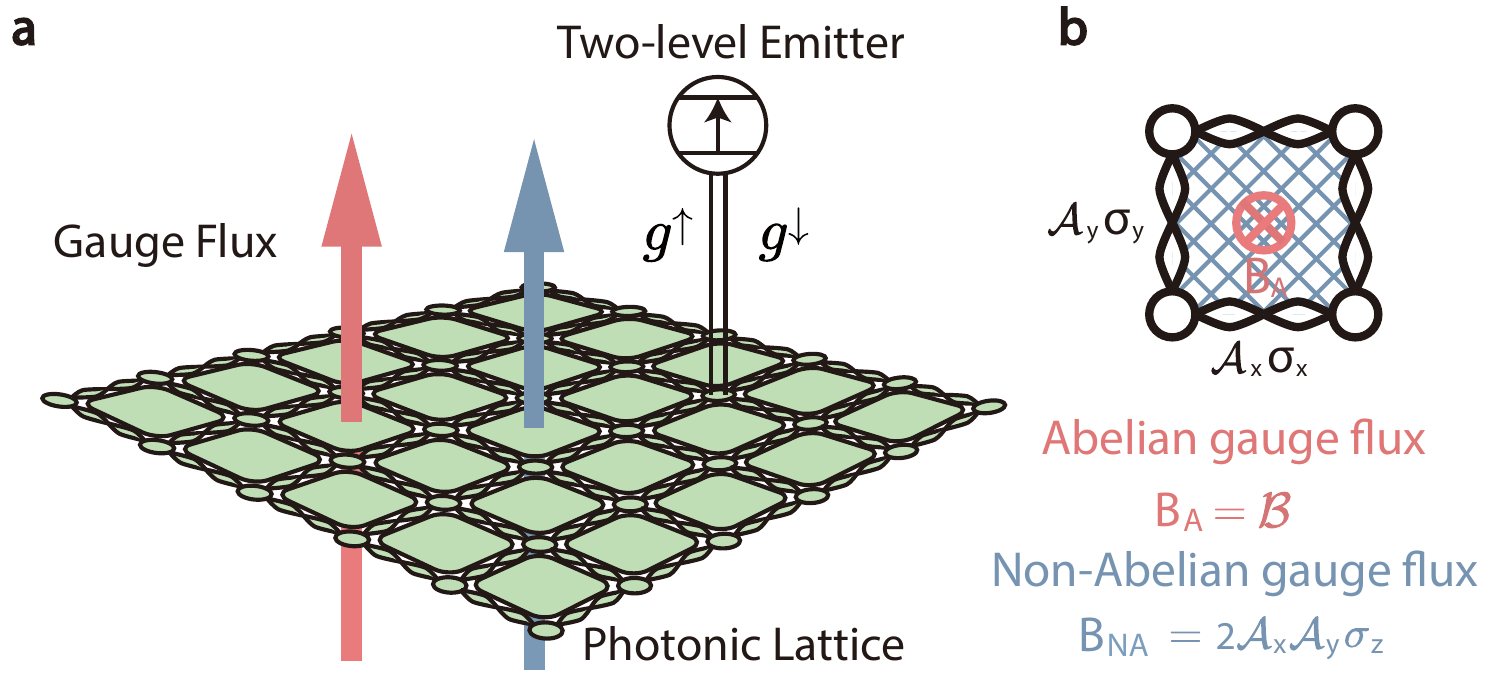}
\caption{\label{fig:Model} 
    \textbf{Interaction between two-level systems and photonic bath of non-Abelian gauge fields.}
    \textbf{a.} Schematic of quantum emitters interacting with photonic lattices threaded by both Abelian (red arrow) and non-Abelian (blue arrow) flux. The coupling strengths between the emitter and photon pseudospins are $g^\uparrow$ and $g^\downarrow$, respectively.
    \textbf{b.} Unit-cell configuration. The red crossed circle and blue grid lines indicate Abelian and non-Abelian gauge flux, respectively.
    }
\end{figure}

Our model setup is shown in Fig.~\ref{fig:Model}a. The system Hamiltonian consists of two-level emitters $H_\text{e}$, photon bath $H_\text{ph}$, and the interaction $H_\text{int}$ between them:
\begin{align}
H=H_\mathrm{e} + H_\text{ph} + H_\text{int}.
\end{align}
We take the ground state energy for the two-level emitters as a reference, \ie $E_g = 0$, and write $H_\mathrm{e}= \sum_i \Delta \ket{e_i}\bra{e_i}$, where $\Delta$ is the detuning between each emitter to the photon lattice. The summation is taken over to account all the emitters. 
The photonic Hamiltonian 
\begin{align}
H_{\text{ph}} = -J \sum_{m, n} & \Psi^\dagger_{m+1,n}e^{-\iu \mathbf{A}_x} \Psi_{m,n}  + \Psi^\dagger_{m,n+1}e^{-\iu \mathbf{A}_y} \Psi_{m,n} + \text{H.c.} \;,
\label{eq:PhotonHamiltonian}
\end{align}
is in the form of a two-dimensional tight-binding square lattice under synthetic gauge fields
$
    \mathbf{A} = \mathbf{A}_\mathrm{A} + \mathbf{A}_\mathrm{NA} = \left(\mathcal{A}_x \sigma_x + \mathcal{B} y /2, \mathcal{A}_y \sigma_y - \mathcal{B} x /2 \right).
    \label{eq:AbelGauge}
$
Here $\mathbf{A}_\mathrm{A}=(\mathcal{B}y/2,-\mathcal{B}x/2)$ describes the U(1) Abelian magnetic field in the symmetric gauge and $\mathbf{A}_\mathrm{NA} = (\mathcal{A}_x \sigma_x,\mathcal{A}_y \sigma_y)$ are the Rashba-type non-Abelian gauge fields, where $\mathcal{A}_{x/y}$ are real numbers.
Other possible choices of SU(2) non-Abelian gauge fields for $H_{\text{ph}}$ are discussed in the Supplementary Material Section 4-5 (SM S4-5).
$\Psi_{\mathbf{r}} = (\Psi_{m,n}^\uparrow, \Psi_{m,n}^\downarrow)^{\mathrm{T}}$ is a spinful field operator that annihilates a photon at site $\mathbf{r}=(m,n)$.
This pseudospin could be a pair of degenerate modes of orthogonal photon circular polarization~\cite{bliokh2015spin, lodahl2017chiral} or the transverse electric and magnetic modes in polaritonic fluids. 
We assume that the emitter couples only to photons at its location. The interaction Hamiltonian can thus be given by
\begin{align}
    H_\text{int} 
        &= \sum_{\mathbf{r}_j} \left( \mathbf{g}_{\mathbf{r}_j}\cdot\Psi_{\mathbf{r}_j}\right)  \sigma^{\mathbf{r}_j}_{+} + \text{H.c.}\;,
    \label{eq:intR}
\end{align}
where $\sigma_\pm$ are the flipping operators on the two-level emitter and the spinful coupling strength $\mathbf{g}_{j}\equiv(g_{\mathbf{r}_j}^{\uparrow},g_{\mathbf{r}_j}^{\downarrow})$, which we call the emitter pseudospin, describes the coupling strengths between the emitter at $\mathbf{r}_j$ and the photon pseudospin components, \ie the overlap between the emitter dipole moment with and the two photon modes~\cite{reitzenstein2010polarization}.
\inset{The amplitude and phase of } $g_{\mathbf{r}_j}^{\uparrow}$ and $g_{\mathbf{r}_j}^{\downarrow}$ can be generically different since the emitter can preferentially couple to one modes over the other.

\section{Non-Abelian Landau dressed states}

\begin{figure}
    \centering
    \includegraphics[width=\linewidth]{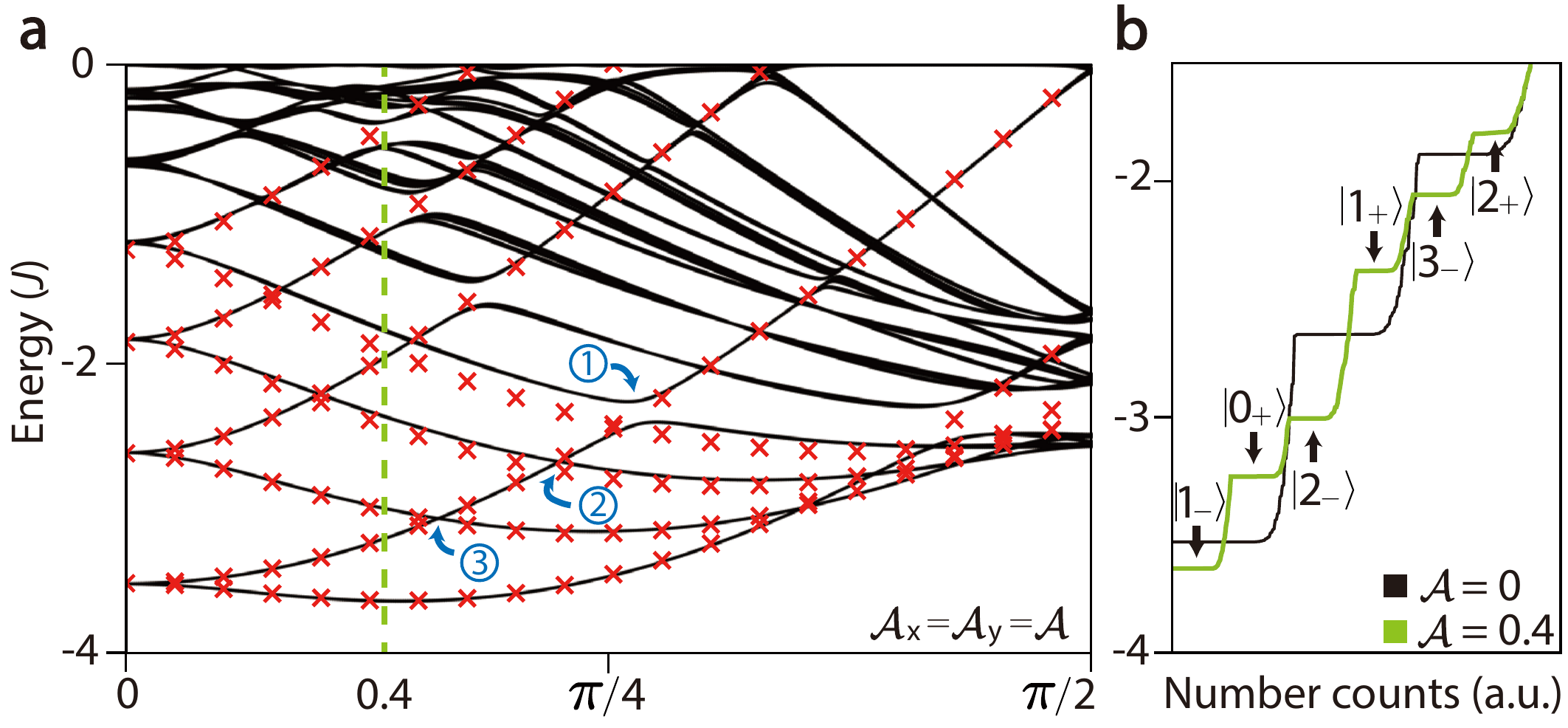}
    \caption{
    \textbf{Non-Abelian Landau dressed states.}
     Spectra of a photonic bath (a) of fixed Abelian magnetic field $\mathcal{B}=0.5$ plotted against the non-Abelian gauge field $\mathcal{A}\in[0,\pi/2]$ and its vertical cuts (b) at $\mathcal{A} =0$ and $\mathcal{A} = 0.4$ (green dashed line), respectively. Black solid lines and crosses label numerical diagonalization and analytical results (for the lowest four Landau levels), respectively. 
    In b
    , each plateau is labeled by the corresponding eigenstates.
    }
    \label{fig:nonAbelianLandauSpectrum}
\end{figure}

We restrict our discussion to the continuum limit of the Abelian gauge field such that the magnetic length $l_B = 1/\sqrt{B}$ is larger than the lattice constant which we set to unity. In this regime, the eigenmodes of the photon lattice are well approximated by Landau orbits. 
This allows us to use the conventional ladder operators for Landau orbits $a = -\iu\left( \mathcal{B} z + 4 \partial_{\bar{z}} \right)/\sqrt{8\mathcal{B}}$, where $z = x-\iu y$, designed for diagonalizing $H_\mathrm{ph}$.
We first focus on the situation of $\mathcal{A}=\mathcal{A}_x=\mathcal{A}_y$, which will be relaxed later.
The study of this particular $H_\mathrm{ph}$ has been performed under $\mathcal{A}\approx \pi/2$ \cite{goldman2009non} and in the continuum limit of non-Abelian fields $\mathcal{A}\ll1$~\cite{burrello2010non,burrello2011ultracold}.
Here, we impose \textit{no} continuum approximations or a particular choice on $\mathcal{A}$ and provide a more generalized solution that continuously interpolates the two previous limiting cases. 
In this photon Hamiltonian, all even orders of the products between $a$ and $a^\dagger$ are diagonal, whereas all their odd orders are anti-diagonal. This is an extension to the quaternions' Euler formula, which we use to separate $H_\mathrm{ph}$ into Abelian and non-Abelian parts $H_\mathrm{ph} \simeq H_{\text{A}} + H_{\text{NA}}$ based on the commutativity of the ladder operators:
\begin{subequations}
    \begin{align}
    H_\text{A} & = J \cos(\mathcal{A}) \left(-4 + 2\mathcal{B} \hat{n} + \mathcal{B} \right),
    \label{eq:AbelPart}
    \\
    H_\text{NA} & = \sqrt{8B} \sin(\mathcal{A}) \left(s_+ a \exp\left(-\frac{\mathcal{B}\hat{n}}{4}\right)  + s_- \exp\left(-\frac{\mathcal{B}\hat{n}}{4}\right) a^\dagger \right).
    \label{eq:nonAbelPart}
    \end{align}
\end{subequations}
where $\hat{n} = a^\dagger a$ is the number operator and the $s_\pm$ are the flipping operators on the photon pseudospins.
\optional{
In Eq.~\eqref{eq:AbelPart}, we only keep the linear order of the number operator, while in Eq.~\eqref{eq:nonAbelPart}, we keep the linear order of the ladder operator in the form of $(a^\dagger a)^p a^\dagger$ ($p$ is an integer), which jointly give rise to the exponential.
}
This Jaynes-Cummings (JC) equivalent form of $H_\text{NA}$ allows one to solve the eigenvalue problem within each $2\times2$ block subspace, consisting of two neighboring Landau levels. The spectra of $H_\text{ph}$ is thus
\begin{align}
\begin{aligned}
E_{l\pm}/J &= (2\mathcal{B}l - 4)\cos(\mathcal{A}) \\
         &\pm \sqrt{\mathcal{B}^2 \cos^2\mathcal{A} + 8\mathcal{B}l\exp\left(-\frac{\mathcal{B}l}{2}\right)\sin^2\mathcal{A}}\;,
\end{aligned}
\label{eq:NonAbelianSpectrum}
\end{align}
with the corresponding eigenstates $|\xi_{l\pm} \rangle$
\begin{subequations}
\begin{align}
    & \hspace{1cm} |{l\pm}\rangle = d_{l\pm}^{\uparrow} |l-1\rangle\otimes\ket{\uparrow} + d_{l\pm}^{\downarrow} |l\rangle\otimes\ket{\downarrow},
    \label{eq:eigenstates}
    \\
    &  d_{l\pm}^{'\uparrow} = - \sqrt{\mathcal{B}} \pm \sqrt{\mathcal{B}+8l\tan^2\mathcal{A}}, \quad
    d_{l\pm}^{'\downarrow} = \sqrt{8l} \tan{\mathcal{A}}\;,    
    \label{eq:NAcoefficients}
\end{align}  
    \label{eq:states_and_coeffs}
\end{subequations}
where $d_{l\pm}^{\uparrow/\downarrow}$ and $d_{l\pm}^{'\uparrow/\downarrow}$ are normalized and unnormalized coefficients of the eigenstates, respectively, $\langle \mathbf{r} |l,m \rangle = \Phi_{lm}(\mathbf{r})$ are Landau orbits in the symmetric gauge, and the angular momentum index $m$ is omitted when irrelevant to the discussion. 
The non-Abelian eigenstates $\ket{l\pm}$ of $H_{\text{A}}+H_{\text{NA}}$ in each subspace are linear combinations of two Landau levels of pseudospin $\ket{\uparrow}$ and $\ket{\downarrow}$ (see SM~ S3).

The direct numerical diagonalization of $H_\mathrm{ph}$ is plotted against these analytical eigenenergies [Eq.~\eqref{eq:NonAbelianSpectrum}] in Fig.~\ref{fig:nonAbelianLandauSpectrum}a, where $H_{A}' = -\frac{\mathcal{B}^2}{8}\cos{\mathcal{A}}(2 \hat{n}^2 + 2 \hat{n} + 1)$, the next leading order correction in $H_\mathrm{A}$, has been included.

Two of its slices at $\mathcal{A}=0$ and $\mathcal{A}=0.4$ are shown in Fig.~\ref{fig:nonAbelianLandauSpectrum}b. 
The Landau levels bifurcate as $\mathcal{A}$ increases as consistent with the appearance of non-Abelian dressed states. 
Excellent agreement is achieved between our analytical solutions and numerics spanning the entire $\mathcal{A}\in[0,\pi/2]$ range.

The crossings and anti-crossings \inset{due to interband coupling} in the spectrum Fig.~\ref{fig:nonAbelianLandauSpectrum}a can now be well explained by our non-perturbative treatment of non-Abelian fields.
In general, the anti-crossings are due to the higher-order $a^p(a^\dagger)^q$ (where integers $p,q$ with $|p-q|>1$ being odd) contributions in the non-Abelian part of $H_\mathrm{ph}$.
A prominent example is the lowest anticrossing near $\mathcal{A}=\pi/4$ (Marker 1 in Fig.~\ref{fig:nonAbelianLandauSpectrum}a), which arises due to the coupling between $\ket{0+}$ and $\ket{4-}$ via $a^3 s_- + (a^\dagger)^3 s_+$. 
In contrast, the $a^p(a^\dagger)^q$ terms where $|p-q|$ is even do not lead to anti-crossing, because they contribute only to the Abelian part of $H_\text{ph}$; an example of this situation is the crossing of $\ket{0+}$ and $\ket{3-}$ (Marker 2 in Fig.~\ref{fig:nonAbelianLandauSpectrum}a).
Moreover, the isotropic non-Abelian fields $\mathcal{A}=\mathcal{A}_x=\mathcal{A}_y$ can also lead to certain crossings, like that between $\ket{0+}$ and $\ket{2-}$ (Marker 3 in Fig.~\ref{fig:nonAbelianLandauSpectrum}a); the anisotropy in $\mathcal{A}$ will introduce an anti-JC coupling that lifts the degeneracy (see SM S5A). 

\section{Nonreciprocity mediated by chiral emitters and SOC photonic bath}

\begin{figure*}
\includegraphics[width=1\linewidth]{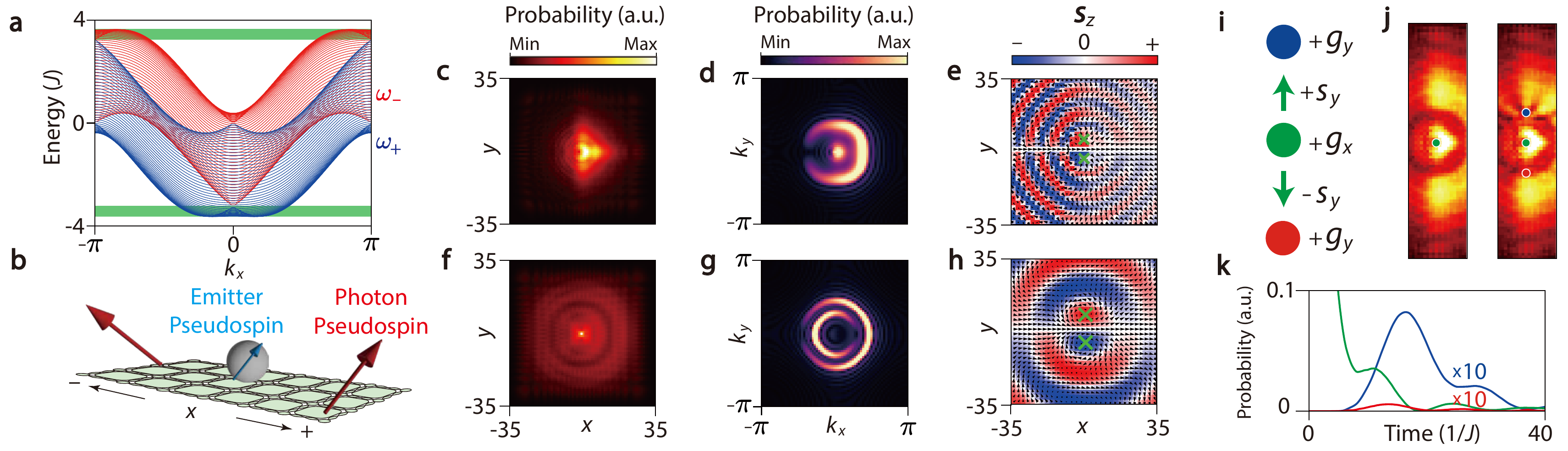}
\caption{\label{fig:PureNonAbelian}
    \textbf{Chiral photon emission and emitter-mediated nonreciprocity with non-Abelian gauge fields.} 
    \textbf{a.} Projected band structures of $\omega_{\pm}(\mathbf{k})$ of the photonic bath. The two green regions mark the detuning windows where the emitter only couples to a single band. $\omega_\pm$ are labeled in a way to be consistent with their corresponding eigenstates $\Psi_{\mathbf{k}\pm}$.
    \textbf{b.} Chiral interaction between the emitter and a SOC photonic band. The emitter pseudospin (blue arrow) becomes parallel and orthogonal to the photon pseudospin (red arrow) at opposite directions, respectively.
    \textbf{c-e.} Chiral emission in real (c) and momentum (d) space and the spin texture (e) when the emitter is detuned into the green region in b. $\mathcal{A}_x = \mathcal{A}_y = \pi/5$ and emitter detuning $\Delta=-3.3J.$ In f, the spin texture $\left(\langle \mathbf{s}_x\rangle,\langle \mathbf{s}_y\rangle\right)$ are drew in black arrows, while $\langle\mathbf{s}_z\rangle$ is color encoded.
    \textbf{f-h.} Same as c-e but for non-chiral emission when the emitter couples to both bands. $\mathcal{A}_x = \mathcal{A}_y = \pi/12$ and emitter detuning $\Delta=-2.7J.$
    The emitter is assumed $+g_x$-oriented with $g_{\uparrow}=g_{\downarrow}=0.2$ throughout c-h.
    Vortices are indicated by green crossings in e and h.
    \textbf{i.} A multi-emitter scattering scenario to show the occurrence of nonreciprocity.
    \textbf{j.} Left panel (zoom-in of c): $y$-symmetric single-emitter emission pattern in the absence of the two scattering emitters; right panel: strong scattering and transparency observed for the identical top and bottom emitters.
    \textbf{k.} Dynamics of the three emitters in the right panel of j. 
    } 
\end{figure*}

After elucidating the photonic eigenstates, we now introduce emitters into our discussion, starting with one emitter located at $\mathbf{r}=0$. 
On the photonic lattice side, we first consider non-Abelian gauge fields only, \ie $\mathcal{B}=0$ (Fig.~\ref{fig:Model}b bottom). It allows convenient diagonalization of $H_\mathrm{ph}$ in momentum space by $\Psi_{\mathbf{r}} = \sum_{\mathbf{k}} \e^{\iu \mathbf{k} \cdot \mathbf{r}} \Psi_{\mathbf{k}}$ in the mixed spin basis $\Psi_{\mathbf{k}\pm} = \left( \Psi_{\mathbf{k}}^{\uparrow} \pm \e^{\iu \theta(\mathbf{k})} \Psi_{\mathbf{k}}^{\downarrow} \right) / \sqrt{2}$ with corresponding eigenvalues $\omega_{\mathbf{k}\pm}$ (Fig.~\ref{fig:PureNonAbelian}a), where $\mathbf{k} = (k_x,k_y)$ and the phase rotation angle $\theta(\mathbf{k}) = \mathrm{Arg}[-\sin(\mathcal{A}_x)\sin(k_x)+\iu\mathrm{sin}(\mathcal{A}_y)\mathrm{sin}(k_y)]$ (see SM S2).
Crucially, $\Psi_{\mathbf{k}\pm}$ constitutes a pair of orthogonal pseudospins at opposite momenta, clearly manifesting SOC.
Moreover, the diagonalized coupling strength $g_{\pm}(\mathbf{k}) = g^{\uparrow}\pm g^{\downarrow} \e^{-\iu\theta(\mathbf{k})}$, associated with eigenstates, $\Psi_{\mathbf{k}\pm}$ are unequal, \ie $g_{+}(\mathbf{k})\neq g_{-}(\mathbf{k})$, and it indicates the chiral coupling $H_{\text{int}}$ between the emitter and the photon bath.
Specifically \inset{for band $\omega_{\mathbf{k}+}$, an $g_x$-oriented emitter is nearly decoupled with the $-s_x$ photons propagating to the $-x$-direction since the emitter and photon pseudospins are orthogonal (Fig.~\ref{fig:PureNonAbelian}b left), whereas it effective couples to $+x$-propagating $+s_x$ photons because their pseudospins are parallel (Fig.~\ref{fig:PureNonAbelian}b right); the opposite is true for band $\omega_{\mathbf{k}-}$. 
}%

The SOC photonic bath $H_{\text{ph}}$ and the chiral coupling $H_{\text{int}}$ jointly enable an emitter-mediated nonreciprocity. 
\inset{Because the two bands exhibit opposite spin-momentum locking, one can achieve strong nonreciprocity by locating the emitter detuning within the SOC band gap (\eg green regions in Fig.~\ref{fig:PureNonAbelian}a) such that the emitter only couples to one of the two bands; this is verified by the directional photon probability distribution (Fig.~\ref{fig:PureNonAbelian}c and d).
On the contrary, when the emitter couples to both of the bands (Fig.~\ref{fig:PureNonAbelian}g), the photon emission distribution is almost symmetric in real space (Fig.~\ref{fig:PureNonAbelian}f).
}

Analytical photon wave function (SM S3) reveals that the nonreciprocity arises jointly from the SOC band splitting and the appearance of vortices in the spin texture: 
\begin{align}
    \begin{pmatrix}
    \phi_{\mathbf{r}}^{\uparrow}
     \\
    \phi_{\mathbf{r}}^{\downarrow}
    \end{pmatrix}
    \appropto
    \sum_{\pm}
    \begin{pmatrix}
        g^\uparrow H_{0}\left(k_{\pm}r \right) \pm \iu \e^{-\iu\theta_{r}} g^{\downarrow}H_{1}\left(k_{\pm}r \right)
        \\
        \pm \iu \e^{\iu \theta_{r}} g^\uparrow H_{1}\left(k_{\pm}r \right) + g^{\downarrow} H_{0}\left(k_{\pm}r \right)
    \end{pmatrix}.
    \label{eq:PhotonRealSpaceAmplitude}
\end{align}
Here \swap{$H_n(r) \equiv H^{(1)}_n(r)$}{$H_n(r)$} is the Hankel function of the first kind. Eq.~\eqref{eq:PhotonRealSpaceAmplitude} is derived under the conditions that (i) the non-Abelian gauge fields are small compared to $\pi/2$ and (ii) the Fermi surfaces are nearly isotropic, i.e. $\omega_{\pm}(\mathbf{k})\approx \omega_{\pm}(|k|)$ and $\theta(\mathbf{k})\approx \theta_{k}$\inset{, and $k_\pm$ are those modes satisfy $\omega_\pm(k_\pm)=\Delta$.}
First, the $\pm$ sign preceding $H_1(k_\pm)$ indicates the opposite spin-momentum locking of the two different bands.
The momentum splitting and SOC band gap increases with the strength of non-Abelian gauge fields, resulting in the eventual decoupling of the emitter with one of the bands under suitable choice of the detuing. Second, note that $H_0(k_{\pm}r)$ and $\iu H_1(k_\pm r)$ both asymptotically approach $\e^{\iu k_\pm r}$ and constructively interfere for $|k_{\pm}r|\gg1$, and the directionality thus hinges on the phase vortex $\e^{\iu\theta_r}$. 
Such a phase vortex indicates that photons carrying orbital angular momentum are excited, as can be seen by the spin textures $\langle \vec{\mathbf{s}} \rangle = \phi_{\mathbf{r}}^{\dagger} \vec{s} \phi_{\mathbf{r}}$ on the Bloch sphere. 
for $+g_x$-oriented emitters, $\left(\langle \mathbf{s}_x\rangle,\langle \mathbf{s}_y\rangle\right)$ adopt the form $(f(r) + \cos(2\theta_r), \sin(2\theta_r))$, where $ f(r) $ depends solely on the distance from the emitter. This configuration generates two vortices equidistant from the emitter (Fig.~\ref{fig:PureNonAbelian}e and h), positioned at the zeros of the Bessel functions in Eq.~\eqref{eq:PhotonRealSpaceAmplitude}.

The nonreciprocity becomes clearly evident in the scattering properties among multiple emitters (Fig.~\ref{fig:PureNonAbelian}i).  Consider a $+g_x$-oriented emitter in the excited state (green circle in Fig.~\ref{fig:PureNonAbelian}i) at the origin $\mathbf{r}=0$ with two $+g_y$-oriented emitters (blue and red circles in Fig.~\ref{fig:PureNonAbelian}i) in the ground state located at $\mathbf{r}=\pm10\hat{\mathbf{y}}$, respectively. 
The detunings of all three emitters are identically chosen such that only one band $\omega_{+}(\mathbf{k})$ needs to be considered. 
In the absence of the top and bottom emitters (Fig.~\ref{fig:PureNonAbelian}j left; a zoom-in of Fig.~\ref{fig:PureNonAbelian}c), the center emitter radiates $\pm s_y$-polarized photons of equal probability distribution toward the $\pm y$ directions, respectively.
Nonetheless, the probability distribution becomes asymmetric as the top and bottom emitters are introduced (Fig.~\ref{fig:PureNonAbelian}j right) and they exhibit distinct scattering dynamics (blue and red curves in Fig.~\ref{fig:PureNonAbelian}k). 
\inset{
Specifically, the $-y$-position emitter (red) exhibits transparency due to its $+g_y$ pseudospin being orthogonal to the downward $-s_y$ emission.
Conversely, the $+y$-positioned emitter (blue) shows strong scattering for its $+g_y$ pseudospin aligning with the upward $+s_y$ emssion.
}
It is emphasized that the ways of spin-momentum locking in the modes are intrinsically specified by the photonic lattice $H_{\text{ph}}$, independent of the choice of the center emitter pseudospin.
Therefore, the contrasting scattering behaviors of the two identical top and bottom emitters against photons moving in opposite $\pm y$ directions manifest nonreciprocity as a result of both synthetic SOC and chiral emitter-photon coupling.

\section{Spin-polarized squeezed Landau polaritons with orbital angular momenta}

\begin{figure*}
\includegraphics[width=\linewidth]{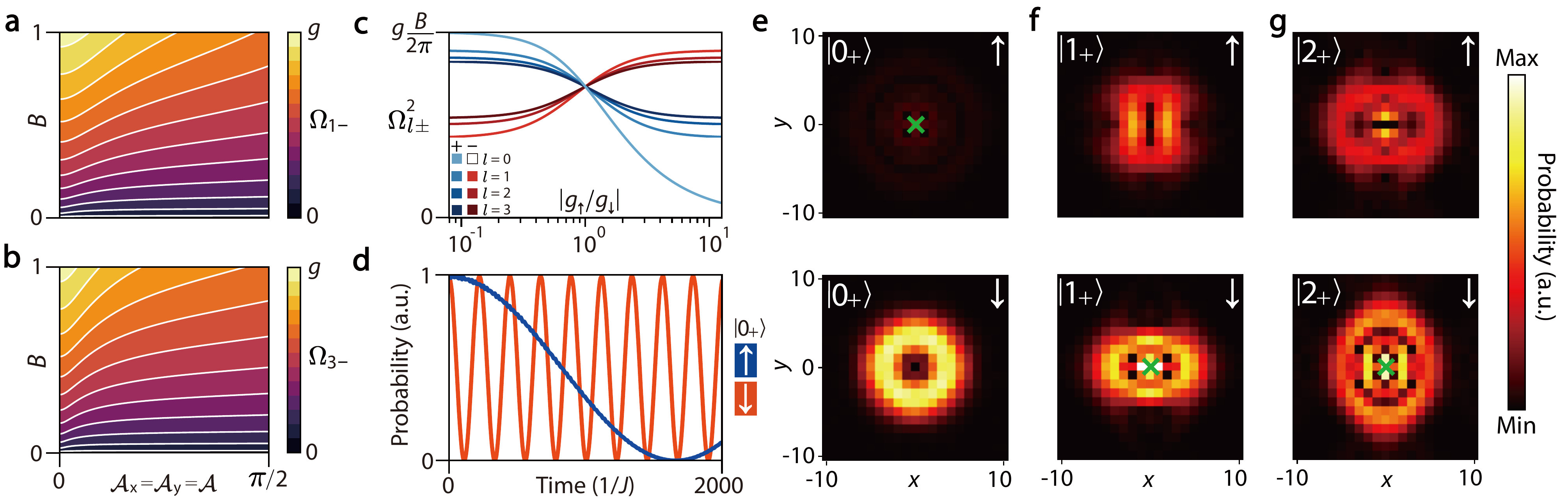}
\caption{
    \label{fig:LLOrbits} \textbf{Spin-polarized squeezed Landau polaritons carrying angular momenta.} 
    \textbf{a-b.} Landau-level--dependent Rabi oscillation.
    Different Rabi frequencies $\Omega_{1-}$ (a) and $\Omega_{3-}$ (b) for a $+g_z$-oriented emitter couples to Landau dressed states $\ket{1-}$ and $\ket{3-}$ respectively.
    \textbf{c.} Rabi frequencies of various dressed Landau polaritons as functions of the contrast of the emitter coupling strength $\abs{g^\uparrow/g^\downarrow}$ under fixed $\mathcal{A}_x=\mathcal{A}_y=0.3$\inset{, with normalization condition $\abs{\mathbf{g}}=g$.}
    \textbf{d.} $+g_z$- (blue) and $-g_z$-oriented (orange) emitter dynamics with $\ket{0+}$ under $\mathcal{A}_x = \mathcal{A}_y = 0.6$.
    \textbf{e-g.} Spin-polarized squeezed Landau polaritons hybridized between a $+g_z$-oriented emitter and $\ket{0+}$ photon under $\mathcal{A}_x = \mathcal{A}_y = 0.6$ (e), a $-g_z$-oriented emitter and $\ket{1+}$ photon under $\mathcal{A}_x=0.4$ and $\mathcal{A}_y = 0.3$ (f), and a $-g_z$-oriented emitter and $\ket{2+}$ photon under $\mathcal{A}_x=0.25$ and $\mathcal{A}_y = 0.15$ (g).
    The green marks indicate the position of the emitter and the pseudospin it couples to. 
    Top and bottom rows correspond to $\ket{\uparrow}$ and $\ket{\downarrow}$ states of the polaritons, respectively.
    In all plots, the magnetic field is $\mathcal{B}=0.5$ and $\abs{\mathbf{g}}=0.05J$.
    }
\end{figure*}

When we turn on the Abelian field with $\mathcal{B}\neq0$, the photon field operators now excite those non-Abelian Landau dressed states. Since the density of states peaks sharply at each Landau level [\onlinecite{ueta1997lattice}], the emitter dynamics become non-Markovian for \textit{any} value of detuning. 
If the emitter's detuning is resonant with a Landau level such that $\Delta = \omega_{l\pm}$, the emitter interacts coherently with that Landau level only, a valid approximation when the light-matter interaction strength is much smaller than the energy difference from the neighboring Landau levels. The emitter's Rabi frequency is thus obtained as (see SM S4)
\begin{align}
    \Omega_{l\pm}^2 = \frac{\mathcal{B}}{2\pi}  \left(\abs{g^{\uparrow} d_{l\pm}^{\uparrow}}^2 + \abs{g^{\downarrow} d_{l\pm}^{\downarrow}}^2\right),
\end{align}
which now depend on the principal quantum number $l$ due to the $l$-dependent $d^{\updownarrow}$. This enables direct identification of emitter-coupled Landau levels through their distinct Rabi frequencies, unlike the uniform Rabi frequencies in pure U(1) Abelian fields. 
Fig.~\ref{fig:LLOrbits}a-b highlight this contrast for a $+g_z$-oriented emitter:
\inset{
When $\mathcal{A}=0$, the non-Abelian dressed states $\ket{1-}$ and $\ket{3-}$ reduce to the Abelian Landau levels $\ket{1}$ and $\ket{3}$} with identical Rabi frequencies; as $\mathcal{A}$ grows, the distinction in their Rabi frequencies becomes more evident.

Aside from the gauge-field tunability in Fig.~\ref{fig:LLOrbits}ab, the non-Abelian Rabi frequencies can also be controlled by the emitter pseudospin. As shown in Fig.~\ref{fig:LLOrbits}c,  \inset{$\Omega_{l\pm}$ become degenerate at $\abs{g^\uparrow}=\abs{g^\downarrow}$ where the $\ket{l+}$ and $\ket{l-}$ states exchange the dominant pseudospin}, but exhibit opposite dependence on the coupling strength contrast $\abs{g^\uparrow/g^\downarrow}$.
Such distinct behavior can be understood from the effective magnetic field $\mathbf{B}_z = 2\mathcal{A}_x\mathcal{A}_y\sigma_z - \mathcal{B}$: within an orthogonal pair of $\ket{l\pm}$, the state with a larger weight on $\ket{\uparrow}$ possesses lower energy. 
The $\ket{0+}$ state deserves special interest. As it lacks the $\left|\uparrow\right\rangle$ component, one might expect a $+g_z$-oriented emitter to exhibit no decay when coupling to \inset{$\ket{0+}$}. However, the emitter can access higher Landau levels via the higher-order coupling $(a^\dagger)^3 s_+ + a^3 s_-$, which leads to an ultra-long Rabi period associated with the $l=3$ states (blue curve in Fig.~\ref{fig:LLOrbits}d).

Still under $\mathcal{A}=\mathcal{A}_x = \mathcal{A}_y$, we obtain the explicit photon wave functions under the emitter excitation \inset{(SM S5)}: 
\begin{align}
    |\Psi^{\updownarrow}_{l\pm}(\mathbf{r},t)|^2 = \bigg| \frac{\sin(\Omega_{l\pm}t)}{\Omega_{l \pm}}
    \sum_{m} & f^{\updownarrow*}_{lm\pm}(\mathbf{r}) \left(g^\uparrow f^{\uparrow}_{lm\pm}(\mathbf{r}_0)+g^\downarrow f^{\downarrow}_{lm\pm}(\mathbf{r}_0)\right)\bigg|^2,
    \label{eq:LandauPhotonDistribution}
\end{align}
where $\mathbf{r}_0$ is the position of the emitter, $f^\uparrow_{lm\pm} = d^{\uparrow}_{l\pm} \Phi_{(l-1)m}$, and $f^\downarrow_{lm\pm} = d^{\downarrow}_{l\pm} \Phi_{lm}$ according to the photonic eigenstates shown in in Eq.~\eqref{eq:states_and_coeffs}. One can perform the summation in Eq.~\eqref{eq:LandauPhotonDistribution} with the help of the identity (SM S5)
\begin{align}
    \sum_m \Phi_{l m}(\mathbf{r}_i) \Phi_{l' m}^*(\mathbf{r}_j) = \sqrt{\frac{\mathcal{B}}{2\pi}} e^{\iu\theta_{ij}} \left\{\begin{matrix} 
    & \Phi_{l\,, l'-l}(\mathbf{r}) \quad (l\leq l')
     \\
     (-)^{l'-l} & \Phi_{l', l'-l}(\mathbf{r}) \quad (l>l')
    \end{matrix}\right.,
    \label{eq:LandauOrbitsSummation}
\end{align}
where $\mathbf{r} = \mathbf{r}_i-\mathbf{r}_j$ and $\theta _{ij} \equiv \frac{\mathcal{B}}{2} \mathbf{\hat{z}} \cdot (\mathbf{r}_i \times \mathbf{r}_j)$. Within the non-Abelian eigenstate, the principal quantum number of Landau orbits on the two pseudospins differs by one, \ie $l-l' = \pm1$. 

Therefore, the dressed Landau photons with angular momenta $m=\pm1$ are excited for $l \neq 0$.
Notably, this selective excitation of the $m=\pm 1$ angular momenta is consistent with their appearances in the photonic wave function under $\mathcal{B}=0$ and $\mathcal{A}\ll1$ [Eq.~\eqref{eq:PhotonRealSpaceAmplitude}].
In contrast, under U(1) Abelian magnetic fields, only photons with zero orbital angular momentum $m = 0$ can exist, as the lack of spin-orbit coupling prevents the excitation of photons with nontrivial orbital angular momentum $m \geq 1$. 
For $l=0$, its spin-down photons can display an angular momentum of $m=3$, shown in Fig.~\ref{fig:LLOrbits}e, because the $|\xi_{3\pm}\rangle$ state is excited (see Fig.~\ref{fig:LLOrbits}d) \inset{via the higher-order coupling} and is further projected on to the unperturbed $|\xi_{0+}\rangle$ state, resulting in such a large angular momentum. 

The physics becomes even richer if the isotropic $\mathcal{A}_x=\mathcal{A}_y$ condition is relaxed, for which we define $q_\pm = (\sin\mathcal{A}_x \pm \sin\mathcal{A}_y)/2$. The non-Abelian part of the Hamiltonian now contains both the JC and anti-JC terms:
$H_\text{NA} = H_{\text{JC}} + H_{\text{AJC}}$, where $H_{\text{JC}} = \sqrt{8\mathcal{B}} q_+ \left(a s_+ + a^\dagger s_-\right)$ and $H_{\text{AJC}} = \sqrt{8\mathcal{B}} q_- \left(a^\dagger s_+ + a s_-\right)$\remove{with no linear order of $\mathcal{A}$ correction on the Abelian part of the photon Hamiltonian}.

\inset{Under a perturbative treatment of the anti-JC term,} the wave functions $f^{\updownarrow*}_{lm\pm}$ now consist of two Landau orbits with a principal quantum number difference of two \inset{(see SM S5A)}.
Using Eq.~\eqref{eq:LandauOrbitsSummation}, the photon wave functions are found to be in the \swap{superpositions}{mixing} of orbital angular momenta of $m=0,\pm2$ on one pseudospin, and states with orbital angular momenta of $m=\pm1,\pm3$ on the other \inset{(see SM S5A)}. 
Thus, within the \textit{same} pseudospin, the distinct $|\Delta m| = 2$ causes interference between different orbital angular momenta channels, giving rise to spin-polarized squeezing of Landau polaritons (see Fig.~\ref{fig:LLOrbits}f-g).
The squeezing direction can be different for different pseudospins, and it depends on both the sign of $q_-$ and the energy levels of the Landau orbits participating in the interference \optional{(see SM S5A)}.
For example, although the signs of $q_-$ for Fig.~\ref{fig:LLOrbits}f and g are the same, the relative energy level positions are switched due to the presence of a crossing in the spectrum at Marker 3 in Fig.~\ref{fig:nonAbelianLandauSpectrum}a, resulting in the opposite squeezing directions in Fig.~\ref{fig:LLOrbits}f and g.

\section{Collective dynamics shaped by nonsymmorphic symmetry}

\begin{figure}
\includegraphics[width=\linewidth]{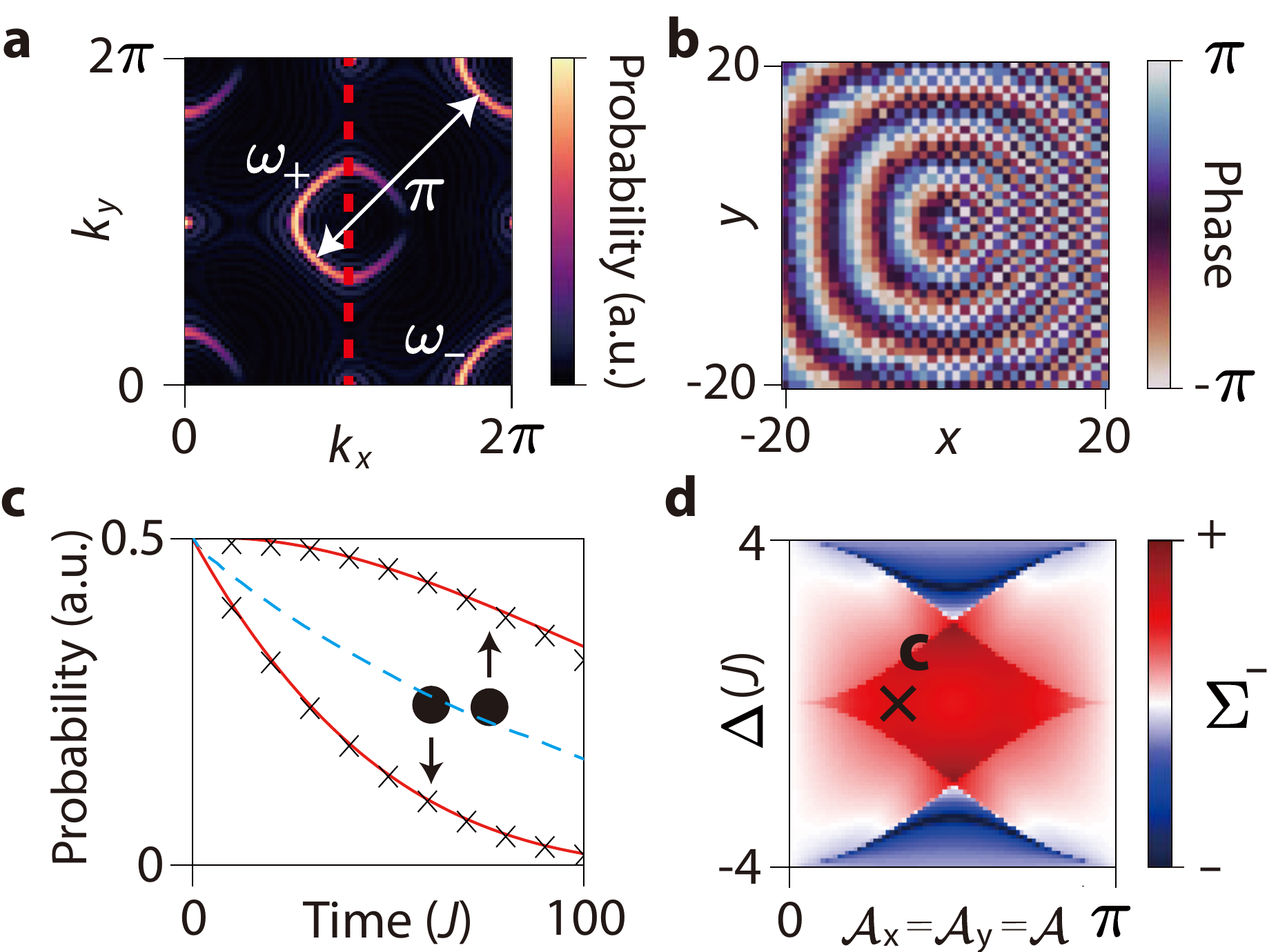}
\caption{\label{fig:TwoEmitters} 
\textbf{Collective dynamics shaped by nonsymmorphic symmetry}. 
    \textbf{a.} Single emitter photon emission wave function in momentum space at zero detuning $\Delta=0$. The double arrow indicates interband mapping from the nonsymmorphic symmetry. 
    \textbf{b.} Real-space phase distribution of a single emitter wavefunction. A staggered $\pi$ phase appears to the right of the emitter.
    \textbf{c.} The symmetry-induced staggered phase results in Purcell enhancement and suppression, respectively, of two identical emitters separated by $\mathbf{r}_{12}=(1,0)$. The dashed line is a single-emitter reference. 
    \textbf{d.} The anti-symmetric part of the cross self-energy $\Sigma^{(-)}$ of the two emitters as a function of non-Abelian gauge fields.
    Here, we choose $\mathcal{A}_x = \mathcal{A}_y = \pi/3$ and emitter detunings $\Delta=0$ with coupling strengths $g^\uparrow = g^\downarrow = 0.1J$.
}
\end{figure}

Next, we turn to the dynamics and interactions of multi-emitter collective systems. Again, we work on pure non-Abelian fields $\mathcal{A}=\mathcal{A}_x=\mathcal{A}_y$ and $\mathcal{B}=0$ and leave the magnetic case $\mathcal{B}\neq0$ in the SM S7. 
The interaction among multiple emitters is mediated by photon exchanges, and, therefore, can be strongly shaped by the crystalline symmetry of the photonic bath. 
We identify a nonsymmorphic symmetry $C(\mathbf{k})=\left[\cos\left(\theta(\mathbf{k})\right), -\iu\sin\left(\theta(\mathbf{k})\right);\iu\sin\left(\theta(\mathbf{k})\right), \cos\left(\theta(\mathbf{k})\right)\right]$ of $H_{\text{ph}}$ such that $\mathcal{C}^\dagger(\mathbf{k}) H(\mathbf{k}+\boldsymbol{\pi})C(\mathbf{k})=-H(\mathbf{k})$ and hence the two bands satisfy $\omega_\pm(\mathbf{k}) = -\omega_{\mp}(\mathbf{k}+\boldsymbol{\pi})$. 
Furthermore, the phase between the photon pseudospin components transforms as $\theta(\mathbf{k}+\boldsymbol{\pi})=\theta(\mathbf{k})+\boldsymbol{\pi}$
between the two bands, where the spins related by the translation $\mathbf{k}\to\mathbf{k}+\boldsymbol{\pi}$ are thus opposite to each other on the Bloch sphere.

Specifically at the vanishing detuning $\Delta=0$, the eigenvalue symmetry reduce to $\omega_\pm(\mathbf{k}) = \omega_\mp(\mathbf{k}+\boldsymbol{\pi})=0$. Hence, as the emitter excites a mode of momentum $\mathbf{k}$ on one band, \eg $\omega_+$, it simultaneously excites a mode of momentum $\mathbf{k}+\boldsymbol{\pi}$ on the other band $\omega_-$. Such correspondence can be seen from the mapping between the two Fermi arcs centered at $(0,0)$ and $(\pi,\pi)$ in the Brillouin zone (Fig.~\ref{fig:TwoEmitters}a). 
On each Fermi arc, the excitation probability again depends on the chiral coupling between the emitter and the spin-momentum locked bands. 
Specifically, a $g_x$-oriented emitter, leads to an overall left propagating wave with momenta centered at $(0,0)$ and an overall right propagating wave with momenta centered at $(\pi,\pi)$ \inset{(see Fig.~\ref{fig:TwoEmitters}a and see SM S6A)}. 
Therefore, as shown in Fig.~\ref{fig:TwoEmitters}b, a striking consequence of the nonsymmorphic symmetry is a staggered $\pi$-phase $\e^{\iu (k_x+\pi) x} = (-1)^{x} \e^{\iu k_x x}$ for right-propagating photons with momentum $k_x+\pi$, whereas this staggered $\pi$-phase is absent for left-moving photons with momentum $k_x$ \inset{(see SM S6A)}, despite that the photon probability distribution $|\phi|^2$ is nonetheless symmetric along the $x$-direction.

Such a symmetry-induced staggered phase substantially modifies the multi-emitter dynamics, the most compelling of which is their distinct behaviors even though identically prepared and immersed in such a homogeneous non-Abelian photonic bath.
Consider two emitters identically prepared in their excited states with a relative distance $\mathbf{r}_{12}=(1,0)$. 
Eq.~\eqref{eq:PhotonRealSpaceAmplitude} indicates that the emitter on the left and right experiences a total field of $\Psi \sim 2 + \e^{\iu\theta_r}\left(\e^{\iu k r} - (-1)^x \e^{\iu k r}\right)$, where $x=1$ and $\theta_r=0$ for the right, and $x=-1$ and $\theta_r=\pi$ for the left emitter, respectively; Thus, constructive and destructive interference can be seen at the location of the left and right emitter, respectively, leading to the Purcell enhancement and suppression of their emission.
However, when the emitters are spaced by even multiples of the lattice period along the $x$-axis (see SM S6B and discussions on other emitter position arrangements), their decay rates do not split. This happens because the staggered phase factor becomes unity, resulting in identical fields experienced by both emitters.

We may moreover define the anti-symmetric part of the cross self-energy $\Sigma^{(-)}\equiv(\Sigma_{12}-\Sigma_{21})/2$ to corroborate the symmetry-shaped dynamics, where $\Sigma_{12}$ and $\Sigma_{21}$ are the collective self-energy functions (see SM S6B). For a pair of identical $g_x$-oriented emitters,
\begin{align}
    \Sigma^{(-)}(z) 
    & = \frac{g^{\uparrow} g^{\downarrow}}{2\pi^2} \int \mathrm{d}^2\mathbf{k} 
    \e^{\iu \mathbf{k}\cdot\mathbf{r}_{12}} \cos \theta(\mathbf{k}) \frac{(\omega_{\mathbf{k+}}-\omega_{\mathbf{k-}})}{(z-\omega_{\mathbf{k+}})(z-\omega_{\mathbf{k-}})}\\
    & \approx 2\pi\iu \cos(\theta_r) \frac{g^{\uparrow} g^{\downarrow} }{N^2} \int \mathrm{d}k
   \frac{kJ_1(k|r_{12}|)(\omega_{\mathbf{k+}}-\omega_{\mathbf{k-}})}{(z-\omega_{\mathbf{k+}})(z-\omega_{\mathbf{k-}})}.
\end{align}
where the second line adopts the approximation of isotropic Fermi arcs and $\theta(\mathbf{k})\approx\theta_k$ as in Eq.~\eqref{eq:PhotonRealSpaceAmplitude} \optional{(see SM S6B)}.
$\Sigma^{(-)}(z)$ is plotted in Fig.~\ref{fig:TwoEmitters}d for $\mathbf{r}_{12}=(1,0)$. We see several requirements for the appearance of asymmetric dynamics from identically prepared emitters.
First, the gauge fields in the photon bath should be non-trivial to achieve SOC splitting between the two bands.
Second, the emitter should simultaneously couple to both photon pseudospins such that $g^\uparrow g^\downarrow \neq 0$, a necessity to achieve chiral emitter-photon interaction.
The self-energy function helps us semi-analytically solves the multi-emitter dynamics within Markovian approximation (\inset{see SM Section 6B}): $|c_{1/2}(t)|^2 = e^{-\Gamma t} \left[1 \pm \sin\delta\sin\left(2|\Sigma_{12}|t\right)\right]$, where $\Gamma$ is the single-emitter decay rate \inset{and $2\delta=\arg\Sigma_{12}-\arg\Sigma_{21}$}. The semi-analytical results are plotted in crossings in Fig.~\ref{fig:TwoEmitters}c \inset{with good agreement achieved with the numerics.}

\section{Summary} 

Our work establishes non-Abelian gauge fields as a transformative paradigm in quantum electrodynamics, revealing their unique capacity to engineer light-matter interactions with topological and symmetry-protected features.
By coupling quantum emitters to photonic bath threaded by non-Abelian flux, we uncovered chiral photon emission with spin-momentum-locked vortices, squeezing and tuning of spin-polarized Landau polaritons, and symmetry-enhanced quantum control of collective emitter dynamics. 
The proposed effects could be realized in state-of-the-art platforms, such as semiconductor quantum dots or solid-state defects coupled to the degeneracy of transverse electric and transverse magnetic bands of exciton polaritons subject to external magnetic fields~\cite{gianfrate2020measurement} and superconducting qubits embedded in topological microwave waveguide networks~\cite{mezzacapo2015non,kim2021quantum}. 
Potential applications of our work include emitter-mediated integrated tunable nonreciprocity, deterministic selection and transfer of quantized angular momenta between light and matter, and quantum control of emitter properties via crystalline symmetries of the bath.


\providecommand{\noopsort}[1]{}\providecommand{\singleletter}[1]{#1}%

\end{document}